\documentstyle{mn}
\def\nl{\\}

\begin{document}

\title[The 3-D structure of the Virgo cluster]
{The 3-D structure of the Virgo cluster from H band Fundamental 
Plane and Tully-Fisher distance determinations.
\thanks{Based on observations taken at TIRGO (Gornergrat, Switzerland),
at the Calar Alto Observatory and at the Observatoire de Haute Provence (CNRS),
(France). 
TIRGO is operated by CAISMI-CNR, Arcetri, Firenze, Italy.
Calar Alto is operated 
by the Max-Planck-Institut f\"ur Astronomie (Heidelberg) jointly with the 
Spanish National Commission for Astronomy.}}

\author[G. Gavazzi et al.]
{G.~Gavazzi,$^1$ A.~Boselli,$^2$
M.~Scodeggio,$^3$  D.~Pierini,$^4$ E.~Belsole$^1$\\
$^1$Universit{\`a} degli Studi di Milano, dipartimento di Fisica, via Celoria, 
16, 20133 Milano, Italy\\
$^2$Laboratoire d' Astronomie Spatiale BP8, Traverse du Syphon, F-13376 
Marseille, France\\
$^3$European Southern Observatory, Karl-Schwarzschild-Str. 2, D-85748
Garching bei M{\"u}nchen, Germany\\
$^4$Max-Planck-Institut f{\"u}r Kernphysik, Postfach 103980, D-69117 Heidelberg, 
Germany \\
}

\date{}
\maketitle


\begin{abstract}

\noindent
We undertook a surface photometry survey of 200 galaxies in the Virgo cluster
(complete to $B<14.0$ mag) carried out in the near-Infrared (NIR) H band.
Combining velocity dispersion measurements from the literature with
new spectroscopic data for 11 galaxies we
derive distances of 59 early type galaxies using the Fundamental Plane (FP) method.
The distance of another 75 late-type galaxies is determined using the Tully-Fisher (TF) method.
For this purpose we use the maximum rotational velocity, as derived from
HI spectra from the literature, complemented with new $H_{\alpha}$ rotation
curves of 8 highly HI deficient galaxies.
The zero-point of the FP and TF template relations are calibrated
assuming the distance modulus of Virgo $\mu_o$=31.0,  
as determined with the Cepheids method. 
Using these 134 distance determinations (with individual uncertainties of 0.35 (TF), 0.45 
(FP) mag)  
we find that the distance of cluster A, associated with M87, is $\mu_o=30.84 \pm0.06$. 
Cluster B, off-set to the south, is found at $\mu_o=31.84 \pm0.10$. This
subcluster is falling onto A at about 750 $km~s^{-1}$. 
Clouds W and M are at twice the distance of A. 
Galaxies on the North-West and South-East of the main cluster A belong to two clouds
composed almost exclusively of spiral galaxies with distances
consistent with A, but with significantly different velocity distributions, 
suggesting that they are falling onto cluster A at approximately 770 $km~s^{-1}$ 
from the far-side and at 200 $km~s^{-1}$ from the near-side respectively.
The mass of Virgo inferred from the peculiar motions induced on its vicinity
is consistent with the virial expectation.
\end{abstract}


\section {Introduction}

Since the pioneering work of de Vaucouleurs (1961) the Virgo 
cluster, the nearest rich galaxy cluster in the northern 
hemisphere, was known to contain significant substructure.
Clouds of galaxies were identified as independent units from 
the main M87 cluster, and they were suggested to lie at larger 
distances, purely on the basis of visual morphological classification. 
The deep photographic survey carried out by Binggeli, Sandage 
and Tammann (1985) with high scale plates, and the spectroscopic
follow-up by Binggeli, Popescu and Tammann (1993) were major 
contributions toward the understanding of the structure of the Virgo 
cluster. These works helped establishing that Cluster A, containing
M87, is separated from cluster B, dominated by M49, and that a number 
of individual clouds, namely cloud M, W, W' and S (using de Vaucouleurs 
designations) are separate entities.\\
Due to their relative proximity, the distance to 5 galaxies in the 
Virgo cluster have been measured with the Cepheids method using
ground-based observations (Pierce et al. 1994) or the HST 
(Ferrarese et al. 1996a; van den Bergh 1996;  Saha et al. 1997). 
The distance of 4 of these (N4321, 4496, 4536, 4571) 
was found consistent with 16 Mpc (corresponding to a distance modulus 
$\mu_o$=31.0), while the remaining one (N4639) was found at a larger
distance of 25 Mpc ($\mu_o$=32.0)(Sandage et al. 1996, Saha et al. 1997).
Distances determinations based on the Tully-Fisher (TF) relation
(Tully \& Fisher 1977) were used by a number of groups to infer the 3-D structure 
of this cluster. Tully \& Shaya (1984) and Pierce \& Tully (1988) 
found evidence for a significant sample depth along the line of sight, 
interpreting their evidence as a presence of infall. 
Fukugita et al. (1993) used their photographic photometry to confirm that 
the cluster has a significant depth. More recently Yasuda et al. (1997) 
re-addressed the issue, and concluded that the B, M and W clouds are at larger 
distance than Virgo itself. Federspiel, Tammann \& Sandage 
(1998) used new B band photometry to pursue the issue. They agree with 
Yasuda et al. (1997) in determining that W and M are distant 
clouds. They confirmed that cluster B is 0.5 mag further 
away than Virgo itself, although the mean recessional velocity of 
this sub-cluster is identical to that of the dominant cluster A 
(Binggeli, Popescu \& Tammann 1993). \nl
Beside the Cepheids and the TF relation, other methods have been used to 
estimate distances to the Virgo spirals. These include the peak of
the luminosity vs. declining ratio of Novae (Ferrarese et al. 1996b),
the expanding photosphere method for the type II supernovae 
(Schmidt et al. 1994), the peak brightness of the Type Ia supernovae 
(Schank 1997) and the bright stars as standard candels (Pierce et al. 1992)
methods. The mean distance modulus derived from these methods is 30.92 $\pm$ 0.28.\nl 
Distance estimates to early-type galaxies have been based on the surface 
brightness fluctuation method (Tonry et al. 1990; Jensen et al. 1996;
Morris \& Shanks 1998; Ajhar et al. 1997),
on Novae luminosities (Della Valle \& Livio 1995), on the globular
clusters luminosity function (Secker \& Harris 1993; Whitmore et
al. 1995; van den Bergh 1996), and on the planetary nebulae
luminosity function (Jacoby et al. 1990; Ciardullo et al. 1998). 
These methods give distances in the range 30.31-32.13. 
M87 itself has been found at $\mu_o$=30.8-31.3. M49 poses a problem, since 
its distance determination ($\mu_o$=31.0) is significantly smaller than 
the distance estimate to cluster B ($\mu_o$=31.8, Federspiel et al. 1998) 
to which it is supposed to belong.\\
Altogether there appears to be unanimous consent that galaxies in 
the Virgo region are not at a unique distance. The W, M 
and B clouds are found further away than A, and evidence for infall 
has been reported. \nl
In this paper we make a further step in the direction of unveiling the 
structure of the Virgo cluster, with two main improvements over 
previous works. Firstly, we combine distances to spirals 
obtained with the TF method with distances to E/S0s obtained with 
the Fundamental Plane (FP) method (Djorgovski \& Davis 1987; 
Dressler et al. 1987), thus increasing the statistical significance of the 
determination. Secondly, for both of these methods we rely on H band 
surface photometry. The advantages of using near-infrared (NIR)
photometry over the B band one are that NIR magnitudes are less 
sensitive to recent episodes of star formation, and they are less 
affected by internal extinction, thus they better trace the luminous 
matter in galaxies (Gavazzi, Pierini \& Boselli 1996b).
Also, spiral disks have a smoother appearance in the NIR, and this
makes it easier to fit photometric profiles to derive the galaxy
total magnitude, and the disk inclination.
Both these factors should contribute to a reduction in the overall
uncertainty with which distance determinations are obtained.\nl
The remainder of this paper is organized as follows. In Section 2 we 
illustrate the sample selected for the present analysis. The H band 
imaging observations and the new spectroscopy are briefly summarized in Section 3. 
The cosmological assumptions, and the derivation of the TF and FP
templates are described in Section 4. Our new distance estimates, 
and a comparison with other estimates from the literature are
presented in Section 5, together with a discussion on the structure
of the Virgo cluster as delineated by these new measurements. 
Our main conclusions are summarized in Section 6.

\section{The Sample}

Galaxies analyzed in this work are selected from the 220 Virgo Cluster
Catalog (VCC; Binggeli, Sandage \& Tammann. 1985) objects with
$B<14.0$.  We add to this sample 13 objects, slightly fainter than this limit, 
that were taken from the CGCG catalogue (Zwicky et al. 1961-68),
because distance estimates are readily available for them. 
For each galaxy a membership estimate is given in the VCC (revised by Binggeli, Popescu 
\& Tammann 1993). Cluster A coincides with the X-ray cluster 
associated with M87. The subcluster B is centered on M49. Other 
aggregates include the M, W and W' clouds and the Southern extension. 
Galaxies with $V_{hel}>3000~km~s^{-1}$ are treated as background
objects, and are excluded from the present analysis.\\
Both the availability of the photometric and spectroscopic data necessary to
build the TF and FP relation, and inclination restrictions that apply
to any TF sample play a role in limiting the number of
galaxies for which we can obtain distance estimates.
We have obtained NIR H or K' band images of 200 out of the 233 
selected galaxies (these observations are
described in Section 3). Of the 89 early-type galaxies imaged, 
48 have central velocity dispersion measurements available in the literature
(McElroy 1995). To these we add 11
new measurements of the central velocity dispersion obtained
at OHP (see Section 3.4). \nl
Of the 111 spiral galaxies imaged, 75 match the criteria for the inclusion in the TF sample, 
i.e.: a) have the correct inclination ($i > 30^{\circ}$) (see Section 4.2); b) have 
high signal-to-noise 21cm HI line profiles (or $H_{\alpha}$ rotation curves)
and, c) meet the additional requirement that the line widths corrected for inclination exceed 
100 $km~s^{-1}$).
The HI line profiles in the literature were individually inspected.
If more than one measurement was available we choosed the one with higher
S/N ratio (references to these measurements are listed in Table 2a).\\
NGC 4496, one of the 5 galaxies with 
Cepheids measurements, is not included in the present analysis because
another galaxy is superimposed on its disk, making the extraction 
of its photometric parameters uncertain.
Conversely two other galaxies with Cepheids (NGC 4321 and 4571)
are included in the present work in spite of their low inclination (27 degrees).


\section{Observations}

\subsection{NIR imaging observations}

In various runs from 1994 to 1997 we took NIR H (1.65 $\mu$m) or 
K' (2.1 $\mu$m) band images of 206 of the 233 galaxies selected from the
VCC and CGCG, using the TIRGO 1.5 m and the Calar Alto 2.2 m telescopes.
Eight galaxies were not observed because they have velocity 
larger than $3000 ~km~s^{-1}$, thus they do not belong to the
cluster, and four because their angular size ($a>4$ arcmin) exceeds 
that of the available NICMOS3 detectors. The remaining 15 objects were
not observed due to time limitations. Images of 6 galaxies are not
usable, because the frames were ruined by stray light from the moon.
Of the 200 images available, 166 were taken in H band, and 34 in K' band.
The K' data are converted into H band using $<H-K'>=0.25 \pm 0.13$ mag. This color,
based on 425 measurements not all included in this work, 
depends little on morphological type (0.24 for E+S0;
0.27 for S+Irr). 

\subsection{Calar Alto Observations of late-type galaxies}

NIR images are available for 111 late-type galaxies: 83 at H band and
28 at K' band. The K' band observations (and 19 of the H band ones)
were taken in 1994-1996 with the Calar Alto 2.2 m telescope, and are 
published in Boselli et al. (1997). H band observations of 15
late-type galaxies were taken in 1995 with the TIRGO 1.5 m telescope, 
and are published in Gavazzi et al. (1996a).\\
Images in the H band for the remaining 49 galaxies
were obtained during three photometric nights of February 1997 
with the Calar Alto 2.2 m telescope, in seeing conditions of typically 
1$\,$-$\,$2 arcsec. The Cassegrain focus of the telescope was equipped
with the MAGIC $256 \times 256$ pixel NICMOS3 infrared array 
(Herbst et al. 1993) with an optical setup of the detector chosen 
to give the largest possible field of view, i.e. 
$6.8 \times 6.8$ arcmin$^2$, with a pixel size of 1.61 arcsec.
The observing technique was identical to that used in previous K' band
observations of late-type galaxies at Calar Alto, as described in 
Boselli et al. (1997). \newline
Galaxies with optical diameter larger than half of the size of the field of
view of the array were observed using a pointing sequence in which 
8 frames are taken, centred on the target, alternated with 8 sky frames, 
positioned along a circular path around the galaxy (off-set by a field of view from
the centre). The 8 on-target fields were dithered by 10 arcsec in 
order to help the elimination of bad pixels. \newline
Galaxies with optical diameter smaller than half of the size of 
the field of view of the array were observed with a pointing sequence
consisting of 9 pointings along a circular path and displaced from 
one-another by 2 arcmin such that the target galaxy is always in the
field. Galaxies with angular sizes larger than the dimension of the 
detector were mapped using pointing sequences expressly prepared 
according to the shape and orientation of the galaxy in the sky, in
order to cover with a mosaic the entire surface of the target. \nl
Typical integration times were of 256-288 seconds. This corresponds 
to the product of the exposure time of the elementary integration 
(1 sec) $\times$ the number of elementary integrations (32) $\times$ the number
of pointings used in each mosaic (8 or 9). \newline
The observations were calibrated and the fluxes transformed into the H band
photometric system using the standard stars in Elias et al. (1982), 
observed hourly throughout the night. The typical uncertainty of the measurements 
is 0.05 mag.\nl
Details on the image analysis and extraction of the photometric parameters 
can also be found in Boselli et al. (1997).
Here it is sufficient to say that the H magnitudes used in the present analysis
were derived by simulating aperture photometry, following the method of
Gavazzi \& Boselli (1996): the counts are integrated in concentric circular rings 
around the galaxy centres to provide curves of growth up to the galaxy
optical diameter (determined in the B at the $25^{th} mag~
arcsec^{-2}$). The H magnitudes used in this work are corrected for internal 
extinction using the prescriptions of Gavazzi \& Boselli (1996).
For 1053 galaxies (not necessarily included in this work)
for which an image is available in H band we
compare the corresponding H magnitudes with those obtained by 
de Vaucouleurs + exponential decomposition of the
ligth profiles obtained along elliptical annuli. We find that on average H
are $0.10 \pm 0.21$ mag fainter than the total (extrapolated to infinity) magnitudes. 

\subsection{TIRGO Observations of early-type galaxies}

NIR images are available for 89 early-type galaxies: 83 at H band, and
6 at K' band. The K' band observations were taken with the Calar Alto
2.2 m telescope, and are published in Boselli et al. (1997). \nl
The H band observations were carried out with the TIRGO 1.5 m
telescope at Gornergrat equipped with the NICMOS3 camera ARNICA 
(Lisi et al. 1993, 1996) in 22 nights from March 13 to April 13, 1997,
as part of an extensive survey of early-type galaxies in the Virgo 
cluster which will be described elsewhere (Gavazzi et al. in preparation).
The optical setting of the camera provides a field of view of 
$4.3 \times 4.3$ arcmin, with 0.96 arcsec pixels. 
The average seeing was 2.3 arcsec.
The observing technique was similar to that used for the Calar Alto
observations, and also in previous observations at TIRGO of late-type 
galaxies, as described in Gavazzi et al. (1996a), and of 
early-type galaxies in the Coma cluster, described in Scodeggio 
et al. (1998a). Here we briefly summarize some of the observing 
parameters relevant to this work. 
All galaxies with apparent B diameter $>1.0$ arcmin were 
observed with pointing sequences which consist of eight frames 
centered on the source, alternated with eight sky frames, positioned 
along a circular path around the source and offset by 4 arcmin. 
The on-source positions were dithered by 10 arcsec to improve the 
flat-fielding, and to facilitate the bad pixel removal. The total 
integration time was 384 seconds, both for the 
target galaxy and for the sky frames.  Galaxies with apparent B 
diameter $< $ 1.0 arcmin were observed with sequences of 9 
pointings along a circular path, displaced from one-another by 1 
arcmin, such that the target galaxy was always in the field. The total 
integration time was 432 seconds. 
The data were calibrated with standard stars in the Elias et al. 
(1982) catalogue, with a typical photometric uncertainty of 0.05 mag.
We checked our photometric calibration against
560 aperture photometry H band measurements available in the literature for
178 galaxies. The agreement was found within 0.10 mag, which we thus take 
as the photometric accuracy of the present investigation.  \nl
The basic image reduction was performed using standard routines 
in the IRAF\footnote{IRAF (Image Reduction and Analysis 
Facility) is distributed by NOAO, which is operated by the 
Association of Universities for Research in Astronomy, Inc. 
(AURA), under cooperative agreement with the National Science 
Foundation.}-STSDAS\footnote{STSDAS(Space Telescope 
Science Data Analysis System) is distributed by STScI, which is 
operated by AURA, under contract to the National Aeronautics and 
Space Administration.}-PROS environment.
The bias-subtracted, flat-fielded, combined, and calibrated images 
were analyzed using the package GAPLHOT (developed for IRAF-
STSDAS mainly by W. Freudling, J. Salzer, and M. Haynes, and 
adapted by one of us (M.S.) to perform the light decomposition of 
early-type galaxies). For each frame the sky background was determined
as the mean number of counts measured in regions of ``empty'' sky, and
it was subtracted from the frame.
Sky-subtracted frames were inspected individually and the light of 
unwanted superposed or nearby stars and galaxies was masked.
The 2-dimensional light distribution of each galaxy was fitted with 
elliptical isophotes, using a modified version of the STSDAS 
${\it isophote}$ package.
The fit maintains as free parameters the ellipse center, ellipticity and 
position angle, and the ellipse semi-major axis is incremented by a
fixed fraction of its value at each step of the fitting procedure. 
Using the fitted parameters  a model of the galaxy light 
distribution is obtained, which is used to compute integrated 
magnitudes as a function of semi-major axis.  \nl
The effective radius $r_e$ and effective surface brightness $\mu_e$ 
(the mean surface brightness within $r_e$) of each galaxy were
obtained by fitting its radial surface 
brightness profile with a de Vaucouleurs $r^{1/4}$ law (de 
Vaucouleurs 1948). The fit was performed from a radius equal to 
twice the seeing radius, out to the outermost isophotes for E 
galaxies; for S0 and S0a galaxies only the central core was fitted. 
The median uncertainty on the determination of log~$r_e$ and 
$\mu_e$ is 0.05 and 0.16 mag., respectively. 

\subsection{Long-slit spectroscopy}

To obtain high dispersion spectra of 21 target galaxies
we used the $\rm 1.93~m$ 
telescope of the Observatoire de Haute Provence (OHP), 
equipped with the Carelec spectrograph (Lemaitre et al. 1990) 
coupled with a 512$\times$512 pixel Tektronix CCD. \newline
The observations were carried out in the 
nights of February 26 - March, 5, 1998 in 2-2.5 arcsec seeing conditions.
The spatial resolution is 1.17 arcsec per pixel. The slit was 1.83 arcsec wide.
We used two grisms, both with a spectral resolution of $\rm 33~\AA$/mm. 
The red grism was chosen 
to give a spectral coverage in the region $\rm 6475-6930~\AA$,
containing the redshifted $\rm H_{\alpha}$ ($\rm \lambda~6562.8~\AA$), 
the [NII] doublet ($\rm \lambda\lambda~6548.1, 6583.4~\AA$) 
and the [SII] doublet ($\rm \lambda\lambda~6717.0, 6731.3~\AA$). This setup
was selected 
to obtain $H_{\alpha}$ rotation curves of 11 spiral galaxies with large ($>0.7$) HI deficiency. 
\newline
Each galaxy was observed with the slit parallel to the major axis, with
an integration time of 15 min, ensuring a signal-to-noise ratio of $\sim$ 10-15
per pixel at $H_{\alpha}$. \nl
For three of these objects (70099, 70115, 70149)
the $\rm H_{\alpha}$ was instead detected in absorption along with the Ca[I] 
($\rm \lambda~6492.5~\AA$), providing us with a useful velocity dispersion measurement.\nl
The blue grism was selected 
to give a spectral coverage in the region $\rm 5050-5505~\AA$,
containing the redshifted Mg[I] triplet ($\rm \lambda\lambda~5167.3-5172.7-5183.6~\AA$)
and the CaFe doublet ($\rm \lambda~5269.0~\AA$). With the blue grism we 
obtained dispersion measurements of 10 early-type galaxies.
Each galaxy was observed with the slit positioned E-W, with
an integration time varying from 20 to 45 min, depending on the brightness of the galaxy,
enough to produce a signal-to-noise ratio of $\sim$ 15 per pixel at Mg[I].
Each observation was preceded by an exposure of a Ar-lamp to ensure the wavelength calibration.
The calibration was checked using known sky lines which were found within 0.1 $\AA$
from their nominal wavelengths. The estimated redshift uncertainty is 10 km $s^{-1}$. \nl
The rotation curves of 8 galaxies were obtained by determining the central wavelength of 
a gaussian fit, pixel per pixel along the slit, to the redshifted $\rm H_{\alpha}$.  
These are given in Fig. 1 and their parameters are listed in Table 1. \newline
The velocity dispersion measurements of 13 galaxies were obtained using the 
Tonry \& Davis (1979) method.
This method is based on a "comparison" between the spectrum of a galaxy whose
velocity dispersion is  to be determined, and a fiducial spectral template 
of a star of appropriate spectral type to contain the wanted absorption lines.
The basic assumption behind this method is that the spectrum of an elliptical
galaxy (and also of the bulge of a disk galaxy) is well approximated
by the spectrum of its most luminous stars (K0-K1 giants), modified only
by the effects of the stellar motions inside the galaxy.
We observed three stars of K type: HD26162, HD132737 and HR5361 with
integration time of 60 sec, to secure a spectrum with sufficient signal-to-noise ratio.
The heliocentric velocities of these stars are known with few  km $s^{-1}$ accuracy.
The quantitative measurements of the spectra have been obtained using the cross-correlation
technique of Tonry \& Davis (1979), implemented in the IRAF task $fxcor$ (see more details
on the application of this technique in Scodeggio 1997).
Since the velocity dispersion inside an elliptical galaxy veries as a function
of radial distance from the center, we extracted from our data
monodimensional spectra of the central 6 arcsec. The measurements
were then corrected to the value that one would obtain at the distance of the 
Coma cluster, using the relation derived by J{\o}rgensen, Franx \& Kj{\ae}rgaard (1996) \nl
$log \sigma/\sigma_{6"}~=~+0.04~log~d_{Virgo}/d_{Coma}$ \nl
where $\sigma_{6"}$ and $\sigma$ are the measured and corrected velocity dispersion,
$d_{Virgo}$ and $d_{Coma}$ are the respective distances, assumed 16 and 88.3 Mpc.
The velocity dispersion measurements obtained in this work are listed
in Column 7 of Table 2b along with their internal errors. The total uncertainty can be obtained
by adding in quadrature the systematic uncertainty of 10 km $s^{-1}$.
Unfortunately two of the newly obtained dispersion measurements (42026 and 70100) are not used
in the present analysis because these galaxies have no H band surface photometry. 

\subsection{The T.F. parameters of HI deficient galaxies}

Some 22 spiral galaxies included in the present analysis have exceedingly large 
($>0.7$) HI deficiency that could produce systematic underestimates of their 
widths, thus of their TF distances.
We inspected these HI profiles individually and we searched in the literature
for alternative maximum rotational velocity measurements, either from CO
observations (Boselli et al. 1995; 
Thronson et al. 1989; Kenney et al. 1995; Kenney \& Young 1988), 
or from $H_{\alpha}$ rotation curves (Chincarini \& de Souza 1985; 
Speraindio et al. 1995; this work).\\
Table 1 reports detailed information on the rotational properties of these galaxies,
as follows:
\nl
Columns 1-3: CGCG, NGC, VCC designations.\nl
Column 4: morphological type as given in the VCC.\nl
Column 5: HI deficiency parameter as defined by Haynes \& Giovanelli (1984).\nl
Column 6: logarithm of the maximum rotational velocity, 
corrected for inclination, as derived from HI observations (average of 20\% and 50\% of the 
peak value). \nl
Column 7: a quality mark given to each HI profile after individual 
inspection: 1 are two-horn, high S/N profiles, 2 are one-horn, high 
S/N profiles, 3-4 are profiles of insufficient quality for T.F. work. \nl
Column 8: galaxy inclination in the plane of the sky (determined
following Haynes \& Giovanelli 1984).\nl
Column 9: reference to the HI data.\nl
Column 10-12: logarithm of the maximum rotational velocity, 
corrected for inclination, as derived from CO observations or $H_{\alpha}$ rotation curves,
with references. \nl
If the three available measurements are found in agreeement we use the HI maximum line widths.
Alternatively we use combinations of CO and $H_{\alpha}$ measurements (when both
are available, as for 99054 and 70097) or we adopt the $H_{\alpha}$ values alone, 
if the data extend sufficiently to ensure that the plateau of the rotation curve is reached. \nl
For three galaxies (70048, 70090 and 70213) observed in this work the $H_{\alpha}$ 
data do not trace the rotation curve
far enough to see the classical S shape, indicative that the plateau of the rotation 
curve has been reached. Their $H_{\alpha}$ maximum rotational velocity 
is smaller than that obtained from the HI line (although of poor quality).
These galaxies have been omitted from the analysis. \newline
One object (70104) has a rotaton curve with marginal change of concavity. However
its maximum velocity determination exceeds that available from the HI profile.
We decided to rely on the $H_{\alpha}$ measurement. The remaning 4 objects observed 
at OHP show a clear S shaped rotation curve. For these we adopt the maximum velocity
obtained from our $H_{\alpha}$ measurement.

\section{The method}

We assume that the average distance of the Virgo cluster A (the membership
is taken from Binggeli, Popescu and Tammann 1993) 
is 16 Mpc ($\mu_o$=31.02), as derived from recent primary distance 
determinations of 4 HST galaxies with Cepheids (van 
den Bergh 1996). We also assume that the difference of distance 
moduli between Virgo cluster A and Coma is 3.71 mag; thus 
$\mu_{o~coma}$=34.73 or $D_{coma}$=88.3 Mpc (van den Bergh 1996). \nl
Given that the average recessional velocity of Coma corrected for the motion 
with respect to the Cosmic Microwave Background (CMB) is $<V>_{CMB}$  = 7185 
$km~s^{-1}$ (Giovanelli et al. 1997), the previous assumptions imply 
$H_o$=81.35 $km~s^{-1}~Mpc^{-1}$. \nl
The infall velocity of the Local Group (LG) toward Virgo is taken to
be 220 $km~s^{-1}$ (see Federspiel et al. 1998); velocities with respect to LG centroid,
$V_{LG}$, are thus computed as $V_{hel}$+220. \nl
Distances to Virgo cluster early- and late-type galaxies are computed
using the FP and TF relation, respectively. Templates for these
relations are obtained from other well studied clusters, scaled according 
to the relative distances on the basis of the above assumptions.

\subsection{The Fundamental Plane relation}

An H band FP template was derived from a sample of 74 galaxies 
in the Coma cluster by Scodeggio et al. (1998a). The best fit to 
this relation, obtained assuming that Coma is at rest in the CMB
reference frame (see, for example, Giovanelli et al. 1997, Scodeggio
et al. 1997a), is: 
$logR_e=-8.354+1.52 log \sigma+0.32 \mu_e$ 
(with the zero-point adapted to $H_o$=81.35). 
The scatter of the template relation, thus the
uncertainty on the distance modulus determination of a single 
galaxy is 0.45 mag. \nl
The fit is obtained minimizing the weighted sum of the orthogonal
distances of the data points from the plane. This is a generalization 
to 3 dimensions of the maximum likelihood method of Press et
al. (1992) (their ``fitexy'' routine), with a modification 
introduced to take into account the high degree of covariance shown
by the uncertainties on the determination of log $R_e$ and 
$\mu_e$ (see Scodeggio 1997 for details).
Uncertainties on the FP parameters are determined using the 
statistical jackknife: N sub-samples, each one composed of N--1 
data-points, are extracted from the original sample of N data-points, 
rejecting in turn one of the data-points. 
The distribution of a certain statistical parameter among those 
N sub-samples is then used to estimate the uncertainty in the value 
of that same parameter, without having to assume an a-priori 
statistical distribution for the parent population of the data-set 
under examination (see for example Tukey 1958, and Efron 1987). \nl 
The FP relation of early-type Virgo galaxies is given in Fig. 2a 
superposed to the fit of the Coma template relation.
The points are coded according to wether $\sigma$ is larger or smaller
than 100 $km~s^{-1}$. It is known that uncertainties in the measurement of such 
low velocity dispersions are larger than in the case of galaxies with
higher velocity dispersion (e.g. Scodeggio et al. 1998b). In fact the dispersion
of Virgo galaxies around the template relation is 0.75 mag if $\sigma<100$ are
excluded, significantly larger than that of the template relation. 
If these 12 galaxies are included, the dispersion rises to 0.85 mag,
enough to suggest to drop galaxies with $\sigma<100$ from the following analysis.

\subsection{The Tully-Fisher relation}

The TF template was derived combining 73 galaxies from the 
clusters A262, Cancer, Coma and A1367. The clusters were considered to
be at rest in the CMB reference frame.
The galaxies were selected from the CGCG (Zwicky et al. 1961-68) (thus 
with $m_p\leq 15.7$)
according to slightly more restrictive criteria than for the Virgo galaxies.
H band data for these galaxies were obtained from Gavazzi \& Boselli (1996)
while the 21cm HI line data were collected from a large number
of sources. It is important to stress here that the method used for determining
the H magnitudes of the template galaxies is identical to that used for the Virgo
galaxies (see Section 3.2).
Galaxies with $DefHI>0.5$ were not used, because the HI line-width in this case
could underestimate the galaxy true rotation velocity.
Also galaxies with $logWc<2.2$ were not used, 
to avoid objects whose correction for turbulent motion is relevant.
We assume that the uncertainty on the measurements of the line width is 10 
$km~s^{-1}$ and that on H band magnitudes is of 0.15 mag. 
The TF relation for the 73 galaxies used in the derivation of the
template is shown in Fig. 2c. The best fitting template, obtained
using the bivariate method (Giovanelli et al. 1997) is given by: 
$H=-2.60 - 7.85~Log W_c$. 
The scatter of the TF template, thus the uncertainty in the distance
modulus of the individual galaxies is 0.35 mag. \nl
The the zero point and slope of the TF template relation are found consistent with the 
parameters of the local TF calibrators. \nl
The TF relation of Virgo late-type galaxies is given in Fig 2b, with 
the template relation superposed. 
The scatter of Virgo galaxies is 0.70 mag, 
thus significantly larger than that of the template relation.
We have checked if to this larger scatter might contribute a not enough
conservative threshold on galaxy inclination ($i>30~deg$). However we found that
for galaxies with inclination $i>45~deg$ the scatter remains 0.70 mag.
Meanwhile we determined that the scatter is not affected by the inclusion of galaxies with
HI deficiency larger than 0.7.
In Figs 2a and 2c all Virgo cluster galaxies are assumed to have a
distance modulus equal to the average distance modulus of Virgo
cluster A ($\mu_o$=31.02). Templates shifted by $\pm$ 1 mag are also given. 
The deviations of the individual objects from the template relations 
can be converted into distances to the individual galaxies.

\section{Results}

\subsection{Derived distances}

The distances to 134 galaxies in the Virgo cluster are listed in Tab. 
2a and b. \nl
Tab. 2a lists the TF parameters of 75 late-type galaxies as follows:\nl
Columns 1-3: CGCG, NGC, VCC designations. \nl
Column 4: morphological type as given in the VCC. \nl
Column 5: membership according to the VCC (revised by    
Binggeli, Popescu and Tammann 1993). \nl
Column 6: HI deficiency parameter as defined by Haynes \& Giovanelli (1984). \nl
Column 7: total H band magnitude corrected for internal 
extinction (see Section 3.1). \nl
Column 8: NIR observing run: CA96 refers to Boselli et al. (1997), CA97 are the
observations taken in 1997 at Calar Alto (see Section 3.1), T95 and T96 are found 
in Gavazzi et al. (1996a). \nl
Column 9: heliocentric recessional velocity from the literature. \nl
Column 10: adopted logarithm of the maximum rotational velocity, 
corrected for inclination, as derived from HI observations (average of 20\% and 50\% of the 
peak value) or from CO observations or $H_{\alpha}$ rotation curves (see Tab. 1). \nl
Column 11: a quality mark given to each HI profile after individual 
inspection: 1 are two-horn, high S/N profiles, 2 are one-horn, high 
S/N profiles, 5 is for one profile which was not published, but was 
considered of good quality given the high S/N ratio. \nl
Column 12: galaxy inclination in the plane of the sky (determined following Haynes 
\& Giovanelli 1984). \nl
Column 13: reference to the adopted rotational velocity. \nl
Column 14: distance modulus as obtained in this work using the TF method.\nl
Column 15: revised region of membership (see Section 5.3 for definition).\nl
A comment is given in Column 16 for the deficient galaxies whose 
maximum rotational velocity is derived from CO profiles or $H_{\alpha}$ 
rotation curves (see Tab. 1).\nl
\nl
Tab. 2b lists the FP parameters of 59 early-type galaxies as follows:\nl
Columns 1 to 5 contain the same information as Tab. 1a.\nl
Column 6: heliocentric recessional velocity. \nl
Column 7-9: central velocity dispersion with error (given for the OHP measurements only) 
and reference. McElroy (1995) (M95) did a compilation of all dispersion measurements available
at the time. Not only the values are averages of various sources, but they
are also corrected for aperture
according to the prescriptions of J{\o}rgensen, Franx \& Kj{\ae}rgaard (1996).
Our data (T.W.) 
are taken through a 2"x6" aperture, corrected similarly (see Section 3.4). \nl
Columns 10 to 15 contain the H band parameters (see Section 3.3). \nl
Column 10-11: Log of the H band effective radius $r_e$,
corrected for seeing according to the prescriptions of Saglia et al. (1993), 
with uncertainty (in arcsec). \nl
Column 12-13: corrected H band effective surface brightness 
with uncertainty (in $mag~arcsec^{-2}$). The correction
includes the cosmological expansion (1+z)$^4$ and K-correction 
(taken to be proportional to 1+z) terms, and the seeing correction, according 
to Saglia et al. (1993). No galactic absorption correction was applied since 
A$_H$~=~0.085~A$_B$ (Pahre et al. 1995), with A$_B~\leq~0.1$ mag in the direction of 
Virgo. \nl
Column 14: seeing during the H band observations.\nl
Column 15: NIR observing run: T97 refers to the 1997 run at TIRGO (this work), 
9 objects were serendipitously observed at Calar Alto (CA94-96) in K'
band and are found in Boselli et al. (1997); one of the T97 objects
was also serendipitously observed at Calar Alto in 1997 in H band.\nl
Column 16: distance modulus as obtained in this work using the FP method.\nl
Column 17: revised region of membership (see Section 5.3 for definition).\nl

\subsection{Comparison with independent distance estimates}

Because many different methods have been applied to derive distance
estimates to Virgo galaxies, a number of consistency checks can be
made between our TF and FP distance determinations and those 
obtained independently by other groups, either with the same or with 
different methods. Fig 3a shows the comparison of our
own distance estimates with those in the literature, obtained with
methods other than the TF or FP relation: Cepheids, type I and II
supernovae, planetary nebulae luminosity function, novae, globular clusters luminosity
function, and surface brightness fluctuations. The rms scatter
in the distance modulus differences,
when measured for E/S0 galaxies alone, is larger (0.40 mag) than that 
measured for the spiral galaxies (0.33 mag), even if we exclude objects 
with $\sigma<100~km~s^{-1}$. 
This latter comparison is based on distance
estimates obtained using Cepheids and SNI and II methods. Assuming
that these methods have an average intrinsic accuracy of 0.2 mag, this
implies that our TF uncertainty is $\sim0.25$ mag. \\
The comparison for E/S0 galaxies is based on distance estimates
obtained using the following methods: surface 
brightness fluctuation (Tonry et al. 1990; Jensen et al. 1996;
Morris \& Shanks 1998; Ajhar et al. 1997),
Novae luminosities (Della Valle \& Livio 1995), the globular
clusters luminosity function (Secker \& Harris 1993; Whitmore et
al. 1995; van den Bergh 1996), and the planetary nebulae
luminosity function (Jacoby et al. 1990; Ciardullo et al. 1998). 
Assuming once again that these methods have an average intrinsic 
accuracy of 0.2 mag, this implies that our FP accuracy is $\sim0.35$
mag. \\
Fig. 3b compares our TF distances with the recent B band 
TF compilation of Yasuda et al. (1997). The rms scatter is 0.47 mag, 
which would correspond to an accuracy of each separate measurement
(assuming equal contributions to the scatter) of roughly 0.3 mag.
This is in good agreement with the internal accuracy estimate of 
Yasuda et al. (1997), but it is surprisingly larger than the
accuracy estimate derived above from the comparison with non TF-based
distance estimates of S galaxies, which should be contributed to by both 
measurement uncertainties and intrinsic variations in galaxy properties. \\
Unfortunately a similar comparison 
cannot be made with the Federspiel et al. (1998) distances, which 
were not published individually.\nl
Distances obtained in this work with the FP method ($\sigma>100~km~s^{-1}$) are compared in 
Fig. 3c with those obtained in the B band by Dressler et al. (1987) and 
Dressler, (1987) with the Dn-$\sigma$ method.
The agreement is not very satisfactory, with a sistematic offset of 0.10 mag
and an rms scatter of 0.40 mag. Since the accuracy
of the B band Dn-$\sigma$ method is expected to be slightly worse than that 
of the FP method, we conclude that the observational scatter of our FP
distance determination is at most 0.3 mag.\\
A test for the internal consistency between our TF and FP distance 
determinations can be done using the interacting system NGC 4435 (S0) +
NGC 4438 (S). The distance moduli of the two galaxies 
(30.59 and 30.26 respectively) agree within the errors. Similarly,
the pair NGC 4649 (S0) + NGC 4647 (S)
has consistent distance moduli (31.37 and 31.19 respectively).
A test of the reproducibility of our results can be done with
the galaxy NGC 4261, which was observed both at TIRGO in 1997 
and, serendipitously, at Calar Alto in 1997; the difference between
the distance moduli based on the two separate observations is 0.15 mag. \nl
A discrepancy is found between our FP-based determination
of the distance of M87, and other determinations found in the
literature. Our value $\mu_o$=31.62 is larger
than 31.30 (van den Bergh 1996), 31.12 (Whitmore et al. 1995) and 30.79 
(Ciardullo et al. 1998). The discrepancy most likely arises from 
an underestimate of the total H magnitude of M87 on our part,
due the fact that M87 fills the TIRGO frame (approx. 4 arcmin),
thus affecting the estimate of the sky background.
Conversely the distance of M49 found in this work $\mu_o$=31.18 is in
excellent agreement with Tonry et al. (1990) ($\mu_o$=31.18) and consistent
with the most recent determination by Morris \& Shanks (1998) ($\mu_o$=30.97).

\subsection{The 3-D Structure of the Virgo Cluster}

The newly obtained distance moduli are plotted against $V_{LG}$ in
Fig. 4, where small dots mark the early type galaxies (FP) separated
into $\sigma<100~km~s^{-1}$ and $\sigma>100~km~s^{-1}$ and larger 
circles mark the spirals (TF), separated into normal and HI 
deficient. The dashed line represents the Hubble flow, drawn for 
$H_o$=81.35. Distance moduli derived with the TF and FP relation have similar
distributions and mean values. HI deficient objects have TF distances
not different from galaxies with normal HI content.
Conversely the FP plane distances of galaxies with $\sigma<100~km~s^{-1}$
are significantly smaller and with higher rms than those with larger 
dispersion. Notice 
that the 4 galaxies with the smallest distance estimates 
are among the objects with the smallest velocity 
dispersions ($\sigma \sim 50-80$ km s$^{-1}$). 
Accordingly their distance estimate uncertainties must be quite large, 
up to the point that we cannot trust them. As anticipated in Section 4.1, 
these galaxies are thus excluded from the rest of the analysis.\\
Three arbitrary regions are evidenced in the figure: \nl
$\mu_o<31.5$ and $\mu_o>31.5$. The latter is split in two 
velocity regions: above and below 1900 $km~s^{-1}$. The region
delimited by $30<\mu_o<31.5$ contains the majority of galaxies 
at the distance to the Virgo cluster. This region contains the majority of 
HI deficient galaxies.\\
A better insight into the structure of the cluster is obtained with 
Fig. 5, which reports the same data as Fig. 4, but with 
symbols whose size decreases with increasing distance, and 
whose shape refers to 7 regions of the Virgo cluster, represented  
in Fig. 6 with dotted lines (galaxies with $\sigma<100~km~s^{-1}$ are omitted). 
Galaxies are assigned to these regions with a criterion that combines their 
position on the sky with their distance and redshift.\\
Distant ($\mu_o>31.5$) galaxies with $V_{LG}>1900$ belong almost 
exclusively to the regions marked M and W. These correspond to the 
M and W clouds, which thus are found to be in Hubble flow.\\
Distant ($\mu_o>31.5$) objects with $V_{LG}<1900$ mostly belong to the 
region marked B. This corresponds to Cluster B of Binggeli et al. 
(1993), but with a smaller extent than in the original definition. 
The mean distance modulus for this structure is found to be 31.84, in
good agreement with the determination by Federspiel et al. (1998) ($\mu_o=31.8$).
To the East of approx $\alpha=12^h25^m$ (M49) all galaxies
have distances not dissimilar from those of Cluster A ($\mu_o~\sim31$).
For example 4 spiral galaxies (NGC 4519, 4332, and 4535, and IC
3521) which are assigned to Cluster B by Binggeli et al. (1985) 
are not confirmed to lie at significantly larger distance 
than Cluster A, since they are found at $\mu_o$=30.2-31.3. 
For these objects our TF distance determinations 
are in agreement with those of Yasuda et al. (1997). 
Another 3 E/S0 Cluster B candidates according to Binggeli et al.
(NGC 4472 (M49), 4526, and 4570) have $\mu_o<31.2$, as determined
using the FP relation. Our low distance estimate of M49 agrees with
independent estimates obtained using all methods quoted above.
Unfortunately this is the only galaxy with independent distance
estimate among this group of early-type objects.	
We propose that M49 belongs to cloud S. Its redshift ($V_{LG}=1200~km~s^{-1}$) is 400 
$km~s^{-1}$ lower than the mean, but it is well within the 
distribution of $V_{LG}$ in this region. \\
The remaining galaxies have $\mu_o<31.5$. They form the main body of the cluster,
indicated here with region A (M87), N, to the NW, E, to the East and S, to the South.\\
Region A coincides with 
Cluster A and with the X-ray position (see Fig. 7 adapted 
from B{\"o}hringer et al. 1994, reproduced on the same scale as Fig. 6). 
As expected, most HI deficient galaxies belong to this region
(see Fig. 8). Two exceptions are surprisingly found at
the southern edge of the cluster in the region of NGC 4636, a strong and 
extended X-ray source (Trinchieri et al. 1994). We argue that a 
significant quantity of extended gas must be associated with this galaxy. \nl
The distribution of $\mu_o$ in the cluster A itself is centered at $\mu_o=30.84$
\footnote {It is important to stress once more that the the absolute distance of
cluster A is not an independent result of this work, since a distance of
Virgo = 16 Mpc has been assumed to calibrate the zero point of both the TF and
FP template relations} 
with a dispersion of 0.45 mag. This is comparable to the 
nominal uncertainty of the distance determination methods (0.35 and
0.45 mag for TF and FP), thus the depth along the line of sight of 
this aggregate cannot be determined. \\
One of the most interesting results of the present 
analysis is that, among galaxies at 
the main cluster distance, those belonging to clouds N and S show a significant 
velocity segregation.
This is illustrated in Fig. 9. The two rightmost 
panels of this figure give histograms of $V_{LG}$ and of $\mu_o$ 
derived from this work for the 7 considered regions.
While the distances of cluster A and clouds S, N, and E are in agreement 
($\mu_o=30.84-31.23$) the only significantly more distant 
structures are cluster B and clouds W and M. Clusters A and B and
cloud E have similar velocity distributions peaked at the standard 
$V_{LG} \sim 1350~km~s^{-1}$. Clouds W and M have instead a higher velocity. 
Clouds S and N have significantly different 
distributions. The N one contains galaxies with $V_{LG}<1300~km~s^{-1}$,
thus blueshifted with respect to Virgo. Similar evidence was pointed out 
by Hoffmann et al. (1989b) and was extensively analyzed by Tully \& Shaya, (1984) 
to model the infall of galaxies on the Virgo cluster.
On the contrary, cloud S contains mainly redshifted galaxies, with 
$750 < V_{LG} < 2700~km~s^{-1}$.
To check if the latter result is not due to the limited statistics of the 
sample used in this work, and since the velocity distribution 
can be derived from a larger body of velocity measurements than the
one represented in our sample of galaxies with distance estimates, 
we determine the velocity distributions in the 7 studied regions using the whole 
VCC (which contains over 400 galaxies with redshift estimates)
(see left panel of Fig. 9). The difference between regions S and N, 
noticed in our smaller sample with distance estimates, is equally 
present in the larger VCC sample, and we conclude it represents a 
real difference between the two clouds.\\
A summary of the mean velocity, velocity dispersion, and
distance modulus determinations for all seven regions is presented in
Tab. 3, and is also shown in Fig. 10, where the average 
velocities and distance moduli are plotted with error bars indicating 
the statistical uncertainties on the determination of the two
quantities (galaxies with $\sigma<100~km~s^{-1}$ are excluded).
Estimates of cluster A are also given separately for early-type
and late-type galaxies, subdivided into HI deficient and HI normal.
Mean quantities and the associated uncertainties 
were computed using the so-called biweight estimators 
(see Beers et al. 1990 for details), that are known to provide 
a robust parameter estimation for samples covering a wide interval 
in size. Statistical uncertainties on the distance modulus
determinations are obtained adding in quadrature the uncertainty on
the FP template zero-point determination and the uncertainty with
which a given sample distance is determined (taking into account the
dispersion that characterizes the FP relation). These should be
conservative estimates, since the uncertainties associated with the TF
relation, that is used to derive 60\% of the distance estimates,
are somewhat smaller than those associated with the FP.
The last column of Table 3 lists the peculiar velocities of the Virgo clouds
with respect to Cluster A (computed as the difference between the actual velocity of the various
entities (Column 3) and their Hubble flow velocity, minus 156  $km~s^{-1}$
which represents the deviation of cluster A itself from the Hubble flow)
, with statistical uncertainties.\\
It appears that the main Virgo cluster (cluster A), cloud E 
and the background cloud W are in near Hubble flow (the deviation of cloud M 
can be assesed with low significance with only 4 distance measurements). 
Cluster B lies at significantly larger distance 
than predicted from its velocity if it was in Hubble flow. Thus it is 
falling into Virgo from the far-side at $\sim~$-760 $km~s^{-1}$.
The two structures N and S have distances similar to the main cluster. 
However their velocities are smaller and larger respectively than 
those of galaxies in dynamic equilibrium with the main body of the 
Virgo cluster. Cloud N, falling from the far-side and cloud S, falling 
from the near-side, have reached the distance of the main cluster.
We remark that these clouds are composed almost entirely of 
spirals (spiral fraction $\sim$80\%), while in all other regions 
the fraction of these galaxies is $\sim~50\%$. \\
The derived parameters of the individual clouds can be compared with similar
determinations carried out by previous investigators.
Yasuda et al. (1997) found  $30.7 < \mu_o < 31.5$ for Cluster A, $\mu_o \sim 31.7$ for Cluster B
and $32.4 < \mu_o < 33.0$ for Clouds W and M, in good agreement with
ours. However these authors claim evidence of infall of clouds W and M, that we do not
confirm.
Federspiel et al. (1998) found $\mu_o = 31.35$ for Cluster A, larger than our
estimate, and $\mu_o \sim 31.81$ for Cluster B, in excellent agreement with the present
determination.

\subsection{The mass estimate of the Virgo cluster}

Determining the mass of the Virgo cluster is beyond the scope of the present
investigation (see Tully \& Shaya 1984 for a comprehensive analysis). 
However a simple consistency check can be done comparing the
virial mass of the cluster with the mass required to perturb the Hubble flow
in the vicinity of Virgo in order to produce the peculiar velocities
observed in the various Virgo clouds and in the Local Group.\\
The virial mass $M_{vir}$ of the Virgo cluster was estimated by Tully \& Shaya (1984) to lie
at $\sim 10^{15} M\odot$ (assuming $H_o$=80 $km~s^{-1}~Mpc^{-1}$).\\
The mass required to perturb the Hubble flow expansion can be estimated
from $\delta v/v= \Omega^{0.6}/3 \times \delta \rho /\rho$, valid in a spherical
geometry for density enhancements over the mean density of the universe
$\delta \rho / \rho \sim 1$ (linear regime) (Davis \& Peebles 1983). 
For larger density enhancements (quasi-linear regime) we use the
analytic correction scheme proposed by Nusser at al. (1991) 
(see their equation 34).
The deviations from the Hubble flow due to the Virgo cluster at the
distance of the Local Group, of cluster
B, of clouds N and W are respectively: $\delta v/v$ = 220/1354, 762/701, 768/56 and 150/1236.
These correspond to density enhancement
$\delta \rho / \rho \sim$ 1.0, 6.7, 85 and 0.75 respectively, assuming $\Omega$=0.3.
Clearly large correction for non-linear regime are required for clouds N and B.  
Using $\rho=2 \times 10^{-29}$, and given the distance of 15 Mpc from the Local Group 
to Virgo, of 8.6 Mpc from cluster B to A, of 0.7 Mpc from cloud N to A and of 15 Mpc from cloud W to A, 
the resulting mass enhancement $\Delta M$ due to the Virgo cluster ranges between 2.1 and 3.1 $\times 10^{15} M\odot$. 
Not surprisingly, this is a factor of 2-3 larger than the virial
expectation. In fact, a significant fraction of the mass is expected
to lie outside the main cluster, given the Virgo complex structure, one
that is far from being a single, dinamically relaxed entity.

\section{Summary and Conclusions}

We have carried out a H band surface photometry survey of 200 galaxies in the
Virgo cluster brigther than $B<14.0$ mag.
These data, complemented with dynamical measurements taken from the literature
and with new spectroscopic data for 19 objects,
allowed us to derive distances to 134 galaxies in this cluster
with either the TF or the FP methods. The individual distance uncertainties
are estimated $\sim$ 0.35 mag and 0.45 mag, for the TF and FP methods respectively. \\
The Virgo cluster is confirmed to be anything but a virialized cluster. Its main 
potential well (cluster A) coincides in position with M87, as 
marked by the distribution of the extended hot gas. An equal 
mixture of E+S0 and of very HI deficient Spiral galaxies populates 
this cluster. The cluster appears elongated in the N-S direction.  
The accuracy of the present distance determinations is still not sufficient to 
convincingly assess a possible extension in depth.\nl
The second major galaxy aggregate, cluster B, is found 
to have a smaller spatial extension than originally proposed by Binggeli, 
Sandage \& Tammann (1985), it lies one 
magnitude further away than Virgo, falling onto the main 
cluster with -760 $km~s^{-1}$ peculiar velocity. 
M49, projected onto cluster B, is found at significantely smaller
distance than B, thus it is proposed to belong to cloud S. 
It remains unexplored whether the hot gas
found in this region is associated with M49 or with cluster B itself. \nl
Two other clouds, almost exclusively composed of spiral galaxies, are 
falling onto the cluster (they are at the same distance as A): the 
one on the North-West is falling from behind (with a peculiar 
velocity of -768 $km~s^{-1}$), the other, to the South of A, is falling from 
the near-side at about +200 $km~s^{-1}$. None of these galaxies suffers 
from severe HI deficiency, indicating that phenomenon has only 
recently (less than few $\times~10^8$ yrs, the time scale for gas depletion)
taken place (see also Tully \& Shaya 1984).
Two more galaxy clouds, M and W, projected on the Western edge of 
Virgo, have distances consistent with B. W is in near 
Hubble flow, while the velocity of M indicates infall. However
its deviation from the Hubble expansion can be assesed with low statistical significance. \nl
The mass of the Virgo cluster inferred from the peculiar motions induced on its
surroundings (Local Group, clouds N, B and W) ranges between 2 and 3 $\times 10^{15} M\odot$,
a factor of 2-3 larger than the virial expectation.\\
We conclude with a note of caution: due to the large gravitational perturbation
induced by cluster A, 
galaxies with large peculiar velocities (regions B, N and S) 
must be cautiously rejected to obtain an umbiased determination of the Hubble constant 
in the Virgo region. Moreover N4639, whose $\mu_o=32.00$ from the present work coincides
with the Cepheyds distance obtained with the HST, should not be used to represent the distance
of Virgo: in fact it belongs to a region (E), containing other galaxies
with similarly large distances.

\section*{Acknowledgments} 
We thank Bruno Binggeli for sending us the VCC in electronic form.
We are grateful to the TAC of the TIRGO and Calar Alto Observatories for the generous
time allocation to this project.
We thank C. Baffa, A. Borriello,  V. Calamai, B. Catinella, I. Randone, 
P. Ranfagni, M. Sozzi, P. Strambio for assistance during the
observations at TIRGO. A special thank to V. Gavriusev for software
assistance at TIRGO, and to T. Beers for permission to use his ROSTAT fortran
code. We also wish to thank an unknown referee whose criticism helped improving
this work.

\newpage

\setcounter{figure}{0}
\begin{figure}
\caption{The rotation curves of 8 HI deficient galaxies observed at the OHP.
The extent of the horizontal scale is up to the galaxy optical diameter.
Galaxies 70048, 70090 and 70213 are not used in the following analysis.}
\end{figure}

\setcounter{figure}{1}
\begin{figure}
\caption{a): the FP relation for 59 Virgo galaxies (dots) separated into $\sigma<100$ and 
$\sigma>100~km~s^{-1}$. The template relation
obtained on the Coma cluster is given using $\mu_o$=31.0 and $\pm 1$ mag. b): the TF relation 
for 75 Virgo galaxies (dots) separated into $Def_{HI}<0.7$ and 
$Def_{HI}>0.7$. The template relation obtained on 4 clusters is given using 
$\mu_o$=31.0 and $\pm$ 1 mag. 
c): The 73 galaxies in the A262, Cancer, Coma and A1367 clusters used 
to derive a template TF relation.} 
\end{figure}

\setcounter{figure}{2}
\begin{figure}
\caption{a): the comparison between the distance moduli obtained 
in this work with those available
in the literature, obtained with other than Tully-Fisher methods. 
b): the comparison between the distance moduli obtained in this work 
with the H band TF method, with the B band ones obtained
by Yasuda et al. (1997).
c): the comparison between the distance moduli obtained in this work (for $\sigma>100~km~s^{-1}$)
and those obtained with the Dn-$\sigma$ method in the B band.
The dotted line is drawn to indicate the one-to-one relationship.} 
\end{figure}

\setcounter{figure}{3}
\begin{figure}
\caption{The distances obtained in this work are plotted against $V_{LG}$. 
Small dots mark the early type galaxies (FP), separated into $\sigma<100$ and 
$\sigma>100~km~s^{-1}$. Larger circles mark 
the spirals (TF), separated into normal and HI deficient. The broken 
line represents the Hubble flow, for $H_o$=81.35 $km~s^{-1}~Mpc^{-1}$.}
\end{figure}

\setcounter{figure}{4}
\begin{figure}
\caption{Same as Fig. 4 with symbols size decreasing 
with increasing distance and whose shape refers to 7 regions of the 
Virgo cluster, represented in Fig. 6 with dotted lines.}
\end{figure}

\setcounter{figure}{5}
\begin{figure}
\caption{The distribution in celestial coordinates of 134 galaxies with distance estimates 
obtained in this work in the region covered by the VCC (solid line). The 7 regions 
of revised membership are given contoured by broken lines. Symbol sizes decrease 
with increasing distance (as in Fig. 5). The positions of M49 and M87 are marked.}  
\end{figure}

\setcounter{figure}{6}
\begin{figure}
\caption{The ROSAT all sky map of Virgo, adapted from B{\"o}hringer et al. (1994) on the
same scale of Fig. 6.}  
\end{figure}

\setcounter{figure}{7}
\begin{figure}
\caption{The distribution in celestial coordinates of 134 galaxies with distance estimates. Small dots mark the early type galaxies (FP) (for $\sigma>100~km~s^{-1}$) and larger circles mark 
the spirals (TF), separated into normal and HI deficient (as in Fig. 4).The positions of M49 and M87 are marked.}  
\end{figure}

\setcounter{figure}{8}
\begin{figure}
\caption{Histograms of $V_{LG}$ in the 7 cluster regions. a): using 433 velocities in the
VCC; b): using 134 galaxies with distance estimates obtained in this work. c): histograms
of the distance moduli obtained in this work in the 7 regions of revised membership.}  
\end{figure}

\setcounter{figure}{9}
\begin{figure}
\caption{The average distances obtained in this work are 
plotted against the average $V_{LG}$ for each of the 7 regions. 
Error bars indicate one-sigma statistical uncertainties on the measured
quantities. The broken 
line represents the Hubble flow, for $H_o$=81.35 $km~s^{-1}~Mpc^{-1}$.}  
\end{figure}

\end{document}